%% file: main.tex
\documentclass[sigconf, screen]{acmart}

\newcommand{\eg}[0]{\textit{e.g., }}
\newcommand{\ie}[0]{\textit{i.e., }}

\newcommand{\ours}{\texttt{\tool{}}}

\usepackage{comment}
\usepackage{stfloats}
\usepackage{microtype}
\usepackage{graphicx}

\AtBeginDocument{%
  \providecommand\BibTeX{{%
    \normalfont B\kern-0.5em{\scshape i\kern-0.25em b}\kern-0.8em\TeX}}}

\usepackage[autostyle,german=guillemets]{csquotes}
\makeatletter

\usepackage{hyperref,quoting}
\usepackage{url}
\quotingsetup{vskip=0pt,font={itshape,raggedright},rightmargin=0pt}
\usepackage{tikz}
\usepackage{cuted}
\usepackage{capt-of}
\usepackage{subcaption}
\usepackage{fancyvrb}
\usepackage{gensymb}
\usepackage{float}
\usepackage{stfloats}
\usepackage{tikz}
\usepackage{xcolor}

\usepackage{acmart-taps,375}

\aptLtoXcmd{\long\def\darkgreysquare#1{\xbox{aptbox}{\XMLaddatt{style}{background-color:\#A9A9A9 ;color: white;border-radius: 30px; padding:5px}{#1}}}}{}

\newcommand{\tool}[1]{\texttt{LearnMate}}

\copyrightyear{2025}
\acmYear{2025}
\setcopyright{rightsretained}
\acmConference[CHI EA '25]{Extended Abstracts of the CHI Conference on Human Factors in Computing Systems}{April 26-May 1, 2025}{Yokohama, Japan}
\acmBooktitle{Extended Abstracts of the CHI Conference on Human Factors in Computing Systems (CHI EA '25), April 26-May 1, 2025, Yokohama, Japan}\acmDOI{10.1145/3706599.3719857}
\acmISBN{979-8-4007-1395-8/2025/04}

\begin{document}


\title{\texttt{\tool{}}\,\raisebox{-0.2em}{\includegraphics[height=1em]{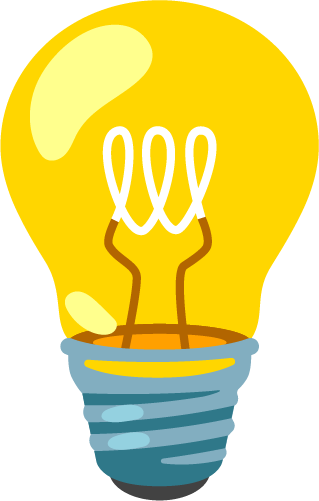}}\,: Enhancing Online Education with LLM-Powered Personalized Learning Plans and Support}








\author{Xinyu Jessica Wang}
\orcid{0009-0002-5519-8432}
\affiliation{%
  \institution{Department of Computer Sciences University of Wisconsin--Madison}
  \country{Madison, Wisconsin, USA}
}
\email{xwang2775@wisc.edu}

\author{Christine P Lee}
\orcid{0000-0003-0991-8072}
\affiliation{%
  \institution{Department of Computer Sciences University of Wisconsin--Madison}
  \country{Madison, Wisconsin, USA}
}
\email{cplee5@cs.wisc.edu}

\author{Bilge Mutlu}
\orcid{0000-0002-9456-1495}
\affiliation{%
  \institution{Department of Computer Sciences University of Wisconsin--Madison}
  \country{Madison, Wisconsin, USA}
}
\email{bilge@cs.wisc.edu}

\renewcommand{\shortauthors}{}

\begin{abstract}

With the increasing prevalence of online learning, adapting education to diverse learner needs remains a persistent challenge. Recent advancements in artificial intelligence (AI), particularly large language models (LLMs), promise powerful tools and capabilities to enhance personalized learning in online educational environments. In this work, we explore how LLMs can improve personalized learning experiences by catering to individual user needs toward enhancing the overall quality of online education. We designed personalization guidelines based on the growing literature on personalized learning to ground LLMs in generating tailored learning plans. To operationalize these guidelines, we implemented \tool{}, an LLM-based system that generates personalized learning plans and provides users with real-time learning support. We discuss the implications and future directions of this work, aiming to move beyond the traditional one-size-fits-all approach by integrating LLM-based personalized support into online learning environments.

\end{abstract}





\begin{teaserfigure}
\end{teaserfigure}
\maketitle


\section{Introduction}

Online learning has become ubiquitous, making online learning courses and platforms accessible to a wide range of learners. Learners include people who lack access to traditional educational tools, children interested in extracurricular activities, and adults who seek continuing education. Massive Open Online Courses (MOOCs) have become particularly popular due to their ability to bring high-quality, low-cost education to large numbers of students \cite{Alseddiqi_AL-Mofleh_Albalooshi_Najam_2023}.



As online learning becomes increasingly accessible to diverse populations, there has been growing emphasis on personalizing learning plans to optimize educational outcomes \cite{basham2016operationalized, bernacki2021systematic, bray2013step}. Personalized learning involves several approaches including (1) technology-based personalization, such as intelligent tutoring systems that adapt to student performance; (2) classroom-based personalization, exemplified by self-paced learning with teacher support; and (3) system-level personalization, which enables multiple pathways for knowledge demonstration \cite{bernacki2021systematic}.

The need for personalized learning is well-documented in existing literature. For example, \citet{bernacki2021systematic} identify several driving factors: the growing diversity of learners; the limitations of traditional ``factory model'' approaches that fail to address individual variability; recent educational policies calling for increased personalization; and the specific needs of students with disabilities whose complex learning profiles demand individualized approaches. However, effectively supporting personalized learning in online learning platforms is still at its early development stage, lacking standardization in implementation and coherent theoretical frameworks \cite{bernacki2021systematic, basham2016operationalized, bray2013step, murtaza2022ai}. Moreover, giving users sufficient control to accommodate their diverse personalized needs and providing accurate and real-time support based on these needs remain open challenges \cite{bernacki2021systematic, zhang2023personal}.



Recent advancements in AI have catalyzed growing interest in leveraging AI for personalized learning experiences \cite{pataranutaporn2021ai, ayeni2024ai, jian2023personalized, davuluri2021ai}. Following the emergence of LLMs, researchers have begun exploring their specific applications in personalized learning and individualized support \cite{park2024empowering, wen2024ai, laak2024ai, nam2024using, hsu2024chime}. Recent LLMs~\cite{gpt-4, geminiteam2023gemini, claude} demonstrate remarkable capabilities across diverse tasks, such as common sense reasoning, adaptation to user-specific needs, and mathematical and visuolinguistic problem-solving \cite{kwon2024toward, krause2023commonsense, lee2025veriplan, didolkar2024metacognitive, cherian2025evaluating}. With carefully crafted prompting strategies, LLMs show particular promise in generating personalized learning plans in online education \cite{ng2024educational, brown2020language, yao2024tree, ma2024beyond}. Additionally, AI-powered conversational services like ChatGPT \cite{gpt-4} have emerged as powerful tools to address persistent challenges in MOOCs, especially in delivering personalized feedback and support to students. Recent studies indicate that these models can effectively facilitate various educational tasks, from providing subject-specific guidance to fostering student engagement \cite{Alseddiqi_AL-Mofleh_Albalooshi_Najam_2023, kayali2023investigation, adetayo2024microsoft}. 

In this paper, we investigate how LLM capabilities can facilitate personalized learning and user support in online educational environments. Our investigation is guided by two research questions: (1) \textit{How should LLMs provide personalized self-paced learning methods and resources based on user needs and preferences;} and (2) \textit{how can LLMs effectively support user learning by offering real-time contextual assistance based on their different needs?}
To address the first research question, we develop personalization guidelines for LLMs to use when generating learning plans, drawing from the growing literature on personalized learning \cite[\textit{e.g.},][]{graham2019k, bernacki2021systematic}. These guidelines focus on four key dimensions: \textit{goals}, \textit{time}, \textit{pace}, and \textit{learning paths}. To operationalize these guidelines and answer the second research question, we present an LLM-based system, called \tool{}, that generates personalized learning plans and provides real-time support to users by monitoring learning progress, tracking study materials, and offering contextual responses to learner queries. Finally, we discuss the potential implications and future directions of this work.

\section{Related Works}
Personalization in online learning continues to grow in popularity, with numerous successful system implementations \cite{bernacki2021systematic, tapalova2022artificial, alamri2021learning}. For example, platforms like Knewton Alta create individualized learning pathways by analyzing student performance and adjusting content delivery in real-time \cite{tapalova2022artificial}. Systems like PLATO and McGraw-Hill's SmartBook demonstrate how adaptive technologies can analyze learning patterns, progress data, and engagement metrics to customize educational experiences, resulting in improved student retention rates from 70.1\% to 89.9\% \cite{alamri2021learning}. 
Despite technological advances, personalized learning faces significant challenges in personalized guidance, real-time support, and companionship for students to maintain engagement in online environments \cite{anshari2016online, garcia2021transformation, murtaza2022ai}. Moreover, giving users enough control to accurately reflect their diverse personalized needs, and for online learning platforms to provide accurate and real-time support based on these needs is often a challenge \cite{bernacki2021systematic}. 

Recent research has explored LLMs for enhancing personalized learning through intelligent tutoring systems. For example, models like PLPP and ALEDU demonstrate significant potential in creating adaptive learning interfaces that generate dynamic, contextually-appropriate educational content \cite{wen2024ai,ng2024educational}. By leveraging natural language understanding and generation capabilities, LLMs can provide personalized, pedagogically sound learning paths tailored to individual student understanding levels \cite{ng2024educational}, improving both user experience and learning outcomes \cite{wen2024ai}.

\section{Design of Personalization Guidelines for LLMs}

\subsection{Personalization Dimensions}

While online learning continues to grow in popularity, significant challenges persist \cite{greenhow2022foundations, garcia2021transformation, anshari2016online, basar2021effectiveness}. As \citet{anshari2016online} highlight, a key question is, \textit{``How can content be tailored for each online learning student based on their Internet behavior?''} This underscores two fundamental challenges in self-directed online learning: the lack of personalized guidance and contextual support.

To address these challenges, we propose a systematic approach that leverages LLMs to generate personalized learning plans that adapt to evolving user conditions. Grounded in principles from the literature on personalized learning, our approach adopts key dimensions that can enable LLMs to collect and incorporate user preferences and needs when developing tailored learning plans.


Our approach adapts the personalized learning framework by \citet{graham2019k}, which includes different dimensions of goals, time, place, pace, and path to support effective personalized learning. Since our focus is on online learning environments that transcend physical limitations---offering greater accessibility and flexibility \cite{anshari2016online}---we have refined this framework to emphasize four key dimensions: goals, time, pace, and path. This adaptation accounts for the inherently location-independent nature of online education while preserving the essential factors that influence learning effectiveness.
Below, we discuss each of the four dimensions adopted for LLMs to generate personalized learning plans.


\begin{enumerate}
        \item{\textbf{Goals:}} Focuses on \textit{what} outcomes to prioritize. This dimension adopts a hybrid goal-setting approach that balances fixed academic requirements with personal learning objectives. It allows learners to pursue individual interests while ensuring alignment with established educational standards. 
        
        \item{\textbf{Time:}} Focuses on \textit{when} instruction takes place and the learner’s \textit{availability}. This dimension incorporates flexible scheduling mechanisms that enable learners to set their time preferences, including weekly time allocations (\eg two to three hours), preferred session formats (\eg one to two hour focused periods vs. 15-30 minute bursts), optimal learning windows, and potential scheduling constraints. 
        
        \item{\textbf{Pace:}} Focuses on \textit{how} quickly learners wish to progress through instruction. This dimension supports structured self-paced learning with customizable progression patterns. Learners can define their preferred learning velocity (\eg steady vs. variable intensity), complexity progression (\eg scaffolded vs. challenge-first), and engagement patterns (\eg consistent effort vs. concentrated bursts). 
        
        \item{\textbf{Path:}} Focuses on \textit{how} learners navigate through learning activities. This dimension enables customization of learning pathways by offering diverse content formats and activities to accommodate different learning styles. Learners can choose from various resource types, such as videos, interactive exercises, and readings, based on their preferences.
\end{enumerate}

These four-dimensional guidelines are designed to help learners accommodate their unique preferences and constraints, supporting their motivation and progress based on individual needs.
These four dimensions are integrated as guidelines for generating personalized learning plans in \tool{}, where the user first defines the \textit{goal} of the learning plan through course selection (detailed in Section \ref{goalGuidelines}), and then defines the \textit{time}, \textit{pace}, and \textit{path} of the learning plan (detailed in Section \ref{personalizationGuidelines}).


\section{User Interaction with \tool{}} \label{sec:tec}
This section introduces the user interaction workflow with \tool{} through a hypothetical scenario and details its technical implementation. We first present the system's three core features, followed by their technical approaches with illustrative examples.
Our implementation uses GPT-4 \cite{gpt-4} for all LLM agents.
Detailed prompts and source code are available in supplementary materials.\footnote{The supplementary materials can be found at \url{https://osf.io/zvh38/?view_only=526aa97c51b442349c77150f95423492}}

\subsection{Guiding User Scenario}

Our hypothetical user scenario features Alex, a marketing professional with a growing interest in space exploration. Despite having limited prior knowledge or experience in astronomy, Alex is eager to understand the foundational concepts of cosmology and astronomy. To achieve this goal, he turns to online learning and decides to try the \textit{Cosmology and Astronomy} course from Khan Academy as his primary resource. The course offers a comprehensive learning experience, including interactive exercises and quizzes on topics such as the Big Bang theory, stellar evolution, and black holes.

\subsection{Course Selection} 
\label{goalGuidelines}   \input{figure_course_selection}

In this step, Alex begins by defining the first dimension of the personalization guideline---his learning \textit{goals}. Upon accessing the system, he can proceed in two ways: (1) specifying his learning interests (depicted as step \darkgreysquare{a} in Figure \ref{fig:courseSelection}) and prompting the system to generate relevant course recommendations with links (step \darkgreysquare{b}), which he can select from directly; or (2) browsing courses by topic (step \darkgreysquare{c}).
Through either approach---specifying interests or browsing topics---Alex can explore the recommended courses and begin building his personalized learning plan.

\paragraph{How it Works}
The system generates personalized course recommendations through a multi-step process driven by user-specified learning goals and language model prompts. When users input their learning goals, the system uses a recommendation mechanism to match them with relevant educational content and links.

Due to limitations in directly accessing Khan Academy's database, we have developed an internal syllabus repository that serves as a proxy for course content. The recommendation process follows two primary pathways: (1) generating tailored course suggestions based on explicit user goals and (2) offering a predefined set of topics with corresponding course recommendations. Once a user selects a course, the system retrieves and presents a detailed syllabus, which serves as the foundation for generating a personalized learning plan. This plan ensures a tailored educational experience that aligns closely with the user's initial learning objectives.

\subsection{Personalized Learning Plan Generation}
\label{personalizationGuidelines}

\input{figure_personalized_plan}

This phase specifies on defining the remaining three dimensions (\ie \textit{time}, \textit{pace}, and \textit{path}) of the personalization guidelines. After selecting a course, Alex configures his learning preferences---time, pace, and style---through the interface (step \darkgreysquare{d}). He selects the corresponding dimension button and specifies his preferences in natural language. Once all preferences are collected, the system generates a text-based personalized learning plan tailored to his inputs (step \darkgreysquare{e}).
        
Based on these preferences, the system now creates a tailored learning plan presented in an interactive calendar, where Alex can easily manage his schedule (step \darkgreysquare{f}). The calendar provides fours views: month, week, day and agenda. He can adjust the timing of modules by dragging and dropping events, can he can also add events or delete event by clicking on the event. These adjustments ensure the plan remains aligned with their learning objectives while accommodating changes in their availability.

\paragraph{How it Works}
The personalized learning plan is generated using two LLM agents: one for strategic plan creation and the other for calendar visualization.
The first agent utilizes language model prompts to generate a structured learning strategy tailored to individual learner needs, following the four-dimensional personalization guidelines. It analyzes key factors such as the natural progression of educational topics, optimal time allocation across subject areas, individual learner constraints, and adaptive learning session strategies. By leveraging these factors, the system ensures a holistic and personalized approach to learning plan generation.

Once the plan is generated, the second agent converts it into a Python \texttt{datetime} format for seamless calendar visualization. Using a template-driven approach with exemplar \texttt{datetime} representations, the agent translates the structured plan into a machine-readable format suitable for calendar integration.
The interactive calendar interface, built with React Big Calendar \cite{reactbigcalendar}, acts as both a visualization tool and an interaction medium, allowing users to flexibly and iteratively refine their learning plans.
        
\subsection{Real-time Learning Support}
\label{RealtimelearningSupport}
\input{figure_learning_support}

Now, Alex can begin his online learning courses with real-time support through \tool{}. Whenever he has questions during his study, he can type them in and receive immediate, context-specific feedback based on the content he has covered (step \darkgreysquare{g}).

\paragraph{How it Works} 
Our initial plan to integrate directly with Khan Academy's API faced constraints due to their stricter embedding policies. Instead, we leveraged YouTube's API \cite{youtube}, which supports Khan Academy's content delivery, enabling a proof-of-concept implementation that meets our system's needs.

The system creates structured YouTube playlists that replicate the course organization of Khan Academy lessons, preserving the original pedagogical sequence. Using the \texttt{youtube\_transcript\_api}, our system translates video content into text transcripts in real time, facilitating comprehensive content analysis and tracking. These transcripts are processed and stored as text files in the database.

When learners pose questions, the system analyzes stored transcripts alongside the user's learning history to generate responses aligned with their learning progression. Each response includes references to relevant video segments and timestamps, allowing learners to easily revisit and reinforce key concepts.
    
\section{Comparison of \tool{} to Single-agent LLM}
In this section, we describe a technical comparison of \tool{} with a single LLM agent. For the monolithic approach, we use a single GPT-4o \cite{gpt-4} agent. We detail our comparative approach below. Specific prompts and records of the comparison can be found in the supplementary materials. 

\subsection{Personalized Learning Framework and Adaptivity}
Our system explicitly incorporates the proposed personalization guidelines, integrating the four dimensions---goals, time, pace, and path---into the planning process. This structured approach provides guidance while empowering students to exercise their agency throughout their educational journey \cite{graham2019k}.
In contrast, the single-agent LLM relies solely on user-provided prompts without a foundational personalization framework, potentially leading to inconsistent outputs and leaving users uncertain about how to effectively control and customize their learning experience.

Additionally, our system performs better compared to the single-agent LLM in handling complex temporal instructions. For instance, we give the same syllabus and asked to implement variable pacing with the prompt \textit{``Allocate more time for fundamental concepts initially, then gradually decrease the duration.''}
In response, the single-agent LLM generates plans with rigid, uniform pacing regardless of the requested temporal variation:
\begin{quote}    
    \textit{Week 1-2:} Unit 1 – Scale of the Universe\\
    \textit{Week 3-4:} Unit 2 – Stars, Black Holes, and Galaxies\\
    \textit{Week 5-6:} Unit 3 – Earth Geological and Climatic History
\end{quote}
In contrast, \tool{} effectively applies dynamic pacing by allocating extended time for foundational topics (five days for Unit $1$) while progressively reducing duration for advanced concepts (four days for Unit $2$, followed by even shorter periods for subsequent units). This approach ensures coherence through structured topic progression and supports user preference for front-loaded learning:
\begin{quote}    
\textbf{Week 1: Unit 1 - Scale of the Universe (5 days)}\\
\textit{Day 1 (Jan 1):} Scale of Earth and Sun (Easy)\\
\textit{Day 2 (Jan 2):} Time Scale of the Cosmos (Medium)\\
\textit{Day 3 (Jan 3):} Light and Fundamental Forces (Medium)\\
\textit{Day 4 (Jan 4):} Special Relativity (Hard)\\
\textit{Day 5 (Jan 5):} Big Bang and Expansion of the Universe (Medium)\\
\textbf{Week 2: Unit 2 - Stars, Black Holes, and Galaxies (4 days)}\\
\textit{Day 6-9 (Jan 6-9):} Life and Death of Stars, Stellar Parallax, etc.
\end{quote}
\tool{} also explicitly labels topic difficulty (easy, medium, hard) to guide resource allocation, a feature absent in the single-agent LLM's plan. See Appendix~\ref{a} for more details.
    
\subsection{Visualization and User Interactivity}
In contrast to the single LLM agent's text-based output, \tool{} integrates an interactive calendar visualization that enables users to both view and manipulate their study plans in real time. This visual-interactive approach enhances plan comprehension and engagement, aligning with research demonstrating how visual representations support effective organization and manipulation of planning activities in educational contexts. Furthermore, \tool{} supports modification through drag-and-drop interactions, providing users with flexible control while maintaining plan coherence. 

    

\subsection{Real-time Learning Assistance and Aligned Contextualization}
Compared to the single-agent LLM's generalized and verbose explanations, our system provides improved real-time and contextual learning assistance by leveraging content-specific APIs and tracking learning history.

The comparison between the single-agent LLM and \tool{} responses reveals several key differences. For example, when asked about wave refraction, the single-agent LLM generates a broad response ($323$ words) with generic physics terminology and examples not covered in the course (\eg water-to-air refraction). In contrast, \tool{} produces a more concise explanation ($243$ words) that explicitly identifies the relevant lesson \textit{Refraction of seismic waves, Cosmology \& Astronomy, Khan Academy} and uses terminology consistent with the course content (\eg "wavefront," "seismic waves").


While the single-agent LLM generates responses based solely on input queries without referencing the learning context, our system leverages a comprehensive learning history stored in transcripts and database records. This enables precise, relevant responses aligned with the user's current progress. For example, \tool{} connects explanations directly to the cosmology curriculum by relating wave refraction to seismic waves and Earth's structure, reinforcing the course's specific learning objectives. In contrast, the single-agent LLM lacks this contextual integration, often resulting in potentially irrelevant responses. Additionally, Our approach further enhances learning efficiency by offering real-time, targeted feedback at any point during the learning journey.


By minimizing verbose and irrelevant content, our system ensures that learners receive concise, actionable support. In contrast, the single-agent LLM's tendency to generate broad, generalized responses can dilute relevance and potentially hinder learning efficiency. See Appendix~\ref{b} for more details.

\section{Discussion and Future Work}

This work explored how LLMs can provide personalized learning experiences that cater to individual user needs, enhancing the overall effectiveness of online education. Many existing online learning platforms offer standardized content and services that do not always align with the diverse learning styles and preferences of users.
We seek to improve users' online learning experiences in two primary ways: (1) enabling LLMs to generate personalized learning plans and (2) utilizing LLMs to provide real-time support during online learning.
To achieve effective personalization, we introduce design guidelines based on four key dimensions---user \textit{goals}, \textit{time}, \textit{pace}, and \textit{path}---to assist LLMs in generating personalized learning plans. These plans are designed to help users tailor online learning services to their unique needs and circumstances, envisioned to address user gaps that may occur with traditional methods.
Additionally, our system leverages LLMs to offer real-time support during learning. Features utilizing LLM capabilities such as instant learning assistance through question answering and interactive companionship can enrich the learning experience, fostering deeper engagement and improved outcomes. Real-time support addresses the lack of immediate assistance and social companionship in online learning \cite{bernacki2021systematic}, which have been shown to reduce motivation and persistence in learners \cite{anshari2016online, garcia2021transformation, murtaza2022ai}.


Looking ahead, our future efforts will focus on enhancing the system's interactivity and usability by introducing intuitive features that align with users' natural learning behaviors.
For example, we plan to expand the system's capabilities to provide additional learning support, such as interactive quizzes, homework assistance, and reflection on the learning material to strengthen social interaction features for improved companionship.
To improve usability, we aim to implement timestamp-based features that guide users to relevant lecture segments based on their questions; generate notes and visual summaries of learning progress; and introduce visual cues to help users navigate content and revisit key concepts efficiently.
Additionally, we will explore methods to strengthen the system's handling of prerequisite relationships between concepts during personalized plan generation. This enhancement will ensure that the system preserves critical knowledge dependencies necessary for building a coherent understanding of complex topics as learning paths are customized to individual preferences, maintaining learning integrity.

Furthermore, we plan to conduct comprehensive evaluations of the system's design and implementation to gain insights into user needs and refine the system accordingly. Our long-term vision includes deploying the system across diverse learning scenarios, such as helping adults acquire new skills, supporting children in remote areas with limited educational resources, and assisting college students in supplementing their coursework. Beyond these scenarios, we envision that the system could also serve as a valuable tool for the general public who rely on platforms like YouTube for self-guided learning.

To maximize accessibility and impact, we aim to develop the system as a flexible plugin that can be integrated with existing online learning platforms such as Khan Academy, MOOCs, and YouTube tutorials. This approach will provide users with personalized learning support within familiar environments, reducing barriers to adoption and ensuring that diverse learners can benefit from tailored educational assistance. Through these efforts, we seek to create a more accessible, adaptive, and user-centered online learning experience.

        


\balance
\bibliographystyle{ACM-Reference-Format}
\bibliography{bibliography}

\appendix
\onecolumn

\section{Comparative Analysis of Personalized Learning Plans} \label{a}

This appendix provides a condensed comparison of personalized learning plans generated by \tool{} and a single-agent LLM. Due to page constraints, only the first portion of each plan is shown. The complete comparison is available in the supplementary materials.

\subsection{User Input}

The user provided the following learning preferences for generating a personalized study plan:

\begin{quote}
\textbf{Time Availability:} ``I can spend two focused hours studying each evening when I feel the most productive. I like to take a break every 40 minutes to stay refreshed.''

\textbf{Pace:} ``I prefer a learning pace that allows me to dedicate more time to mastering fundamental concepts at the start and gradually reduce the intensity as I progress.''

\textbf{Content Preferences:} ``I learn best through engaging and visually rich content, so I prefer video-based lessons that explain complex ideas with animations or diagrams.''
\end{quote}

\subsection{Single-agent LLM Response (First Four Weeks)}

\begin{quote}
\textbf{Weekly Study Schedule Overview}\\
Daily Study Time: 2 focused hours\\
Break Schedule: Study for 40 minutes, take a 10-minute break, then resume.

\textbf{Week 1-2: Unit 1 – Scale of the Universe}\\
Goal: Master the universe's scale and fundamental concepts (\eg light, forces, Big Bang).

Weekdays:
\begin{itemize}
    \item Session 1: Watch video-based lessons for ``Scale of Earth, Sun, Galaxy, and Universe.''
    \item Session 2: Explore interactive tools for visualizing cosmic scales.
    \item Session 3: Study ``Time Scale of the Cosmos'' and ``Light and Fundamental Forces.''
    \item Session 4: Review special relativity (allocate extra time, as it's labeled ``hard'').
\end{itemize}

Weekend:
\begin{itemize}
    \item Take the Vocabulary Quiz.
    \item Begin light reading on ``Big Bang and Expansion of the Universe.''
\end{itemize}

\textbf{Week 3-4: Unit 2 – Stars, Black Holes, and Galaxies}\\
Goal: Build understanding of stellar phenomena and galactic structures.

Weekdays:
\begin{itemize}
    \item Session 1: Focus on ``Life and Death of Stars.'' Use videos with animations showing stellar cycles.
    \item Session 2: Study quasars and galactic collisions—break down complex topics into smaller parts.
    \item Session 3: Dive into visualizations of stellar parallax and Cepheid variables.
\end{itemize}

Weekend:
\begin{itemize}
    \item Complete Vocabulary Quiz.
    \item Practice recalling key concepts through short written summaries.
\end{itemize}

[additional weeks omitted; see supplementary materials for complete plan]
\end{quote}

\subsection{\tool{} Response (First Nine Days)}

\begin{quote}
\textbf{Astronomy Course Schedule - January 2025}

\textbf{Week 1: Unit 1 - Scale of the Universe}

Day 1 (Jan 1)\\
6:00 PM - 6:40 PM: Scale of Earth and Sun (Easy)\\
6:40 PM - 6:50 PM: Break\\
6:50 PM - 7:30 PM: Scale of Galaxy and Universe (Easy)

Day 2 (Jan 2)\\
6:00 PM - 6:40 PM: Time Scale of the Cosmos (Medium)\\
6:40 PM - 6:50 PM: Break\\
6:50 PM - 7:30 PM: Time Scale of the Cosmos (Medium)

Day 3 (Jan 3)\\
6:00 PM - 6:40 PM: Light and Fundamental Forces (Medium)\\
6:40 PM - 6:50 PM: Break\\
6:50 PM - 7:30 PM: Light and Fundamental Forces (Medium)

Day 4 (Jan 4)\\
6:00 PM - 6:40 PM: Special Relativity (Hard)\\
6:40 PM - 6:50 PM: Break\\
6:50 PM - 7:30 PM: Special Relativity (Hard)

Day 5 (Jan 5)\\
6:00 PM - 6:40 PM: Big Bang and Expansion of the Universe (Medium)\\
6:40 PM - 6:50 PM: Break\\
6:50 PM - 7:30 PM: Big Bang and Expansion of the Universe (Medium)

\textbf{Week 2: Unit 2 - Stars, Black Holes, and Galaxies}

Day 6 (Jan 6)\\
6:00 PM - 6:40 PM: Life and Death of Stars (Medium)\\
6:40 PM - 6:50 PM: Break\\
6:50 PM - 7:30 PM: Life and Death of Stars (Medium)

Day 7 (Jan 7)\\
6:00 PM - 6:40 PM: Stellar Parallax (Easy)\\
6:40 PM - 6:50 PM: Break\\
6:50 PM - 7:30 PM: Quasars and Galactic Collisions (Hard)

Day 8 (Jan 8)\\
6:00 PM - 6:40 PM: Quasars and Galactic Collisions (Hard)\\
6:40 PM - 6:50 PM: Break\\
6:50 PM - 7:30 PM: Cepheid Variables (Medium)

Day 9 (Jan 9)\\
6:00 PM - 6:40 PM: Cepheid Variables (Medium)\\
6:40 PM - 6:50 PM: Break\\
6:50 PM - 7:30 PM: Unit 3: Plate Tectonics (Medium)

[additional days omitted; see supplementary materials for complete plan]
\end{quote}

\section{Comparative Analysis of Real-time Learning Support} \label{b}

This appendix provides a side-by-side comparison of outputs from \tool{} and a single-agent LLM when responding to the same user query during a learning session on wave refraction. 

\subsection{User Question During Study}

During a lesson on wave refraction in the Cosmology \& Astronomy course, the user asked:

\begin{quote}
``In the analogy of a car crossing from mud (slow medium) to a road (fast medium), how does the idea of traction—where one wheel moves faster than the others—explain the bending of a wave's direction during refraction? Why does this depend on the wave's angle of approach?''
\end{quote}

\subsection{Single-agent LLM Response}

The single-agent LLM provided the following response:

\begin{quote}
The analogy of a car crossing from mud (a slow medium) to a road (a fast medium) helps illustrate the concept of wave refraction in physics, particularly how waves bend as they move from one medium to another at an angle.

\textbf{Traction Analogy:}

In the car analogy, when a car moves from a slow medium (mud) to a fast medium (road), the wheels in contact with the mud (slow medium) move more slowly than those in contact with the road (fast medium). This difference in speed causes the car to turn or change direction. Similarly, when a wave moves from one medium to another, the part of the wave that enters the faster medium will travel faster than the part that is still in the slower medium. This creates a bending effect in the direction of the wave's motion, known as refraction.

\textbf{Refraction and the Angle of Approach:}

Refraction occurs because the speed of the wave changes when it enters a different medium. The angle of approach determines how much the wave will bend because it affects the difference in the wave's speed between the two mediums:

\begin{itemize}
    \item When the wave enters a faster medium (\eg from water to air), the part of the wave entering the faster medium moves faster than the part still in the slower medium. This causes the wave to bend toward the normal (the imaginary line perpendicular to the boundary between the two media).
    \item When the wave enters a slower medium (\eg from air to water), the opposite happens. The wave slows down in the slower medium, causing the wave to bend away from the normal.
\end{itemize}

The bending depends on the wave's angle of incidence (the angle at which the wave approaches the boundary). If the wave approaches the boundary at a steep angle, there is a larger difference in the speeds between the parts of the wave in each medium, leading to a greater bend. If the wave approaches more perpendicular to the boundary (at a small angle), the difference in speeds is less pronounced, resulting in a smaller change in direction.

\textbf{In summary:}
\begin{itemize}
    \item The change in speed of the wave as it enters a new medium causes parts of the wave to move at different rates, leading to bending (refraction).
    \item The angle of approach affects the degree of refraction: steeper angles lead to more pronounced bending.
\end{itemize}
\end{quote}

\subsection{\tool{} Response}

\tool{} provided the following response:

\begin{quote}
\textbf{Relevant Lesson: Refraction of seismic waves | Cosmology \& Astronomy | Khan Academy}

The analogy of a car crossing from a muddy area to a paved road helps explain the concept of refraction, which is the bending of a wave's direction as it passes from one medium to another with a different speed of wave propagation. In the analogy, the car's wheels represent the wavefront of a seismic wave (or any type of wave, such as light or sound), and the mud and road represent two different media through which the wave travels.

When the car's wheels on one side (say, the right side) hit the road (a faster medium) while the wheels on the other side (the left side) are still in the mud (a slower medium), the right wheels can move faster than the left wheels. This difference in speed causes the car to turn towards the right. Similarly, when a wavefront hits the boundary between two media at an angle, the part of the wavefront that enters the faster medium first speeds up, causing the wave to bend towards the normal (an imaginary line perpendicular to the boundary). This bending is due to the wave's speed changing as it enters the new medium.

The amount of bending depends on the angle at which the wave approaches the boundary. If the wave approaches the boundary at a perpendicular angle (90 degrees to the boundary), there is no bending, and the wave continues straight. However, if the wave approaches at any other angle, the change in speed at the boundary causes the wave to bend.

This concept is crucial in understanding how seismic waves travel through the Earth's interior and how scientists can infer the structure of the Earth's layers. Seismic waves refract, or bend, as they pass through layers of different densities and elastic properties, which helps seismologists determine the composition and properties of these layers.
\end{quote}

\end{document}

%% file: figure_course_selection.tex
\begin{figure*}[!th]
  \includegraphics[width=\textwidth]{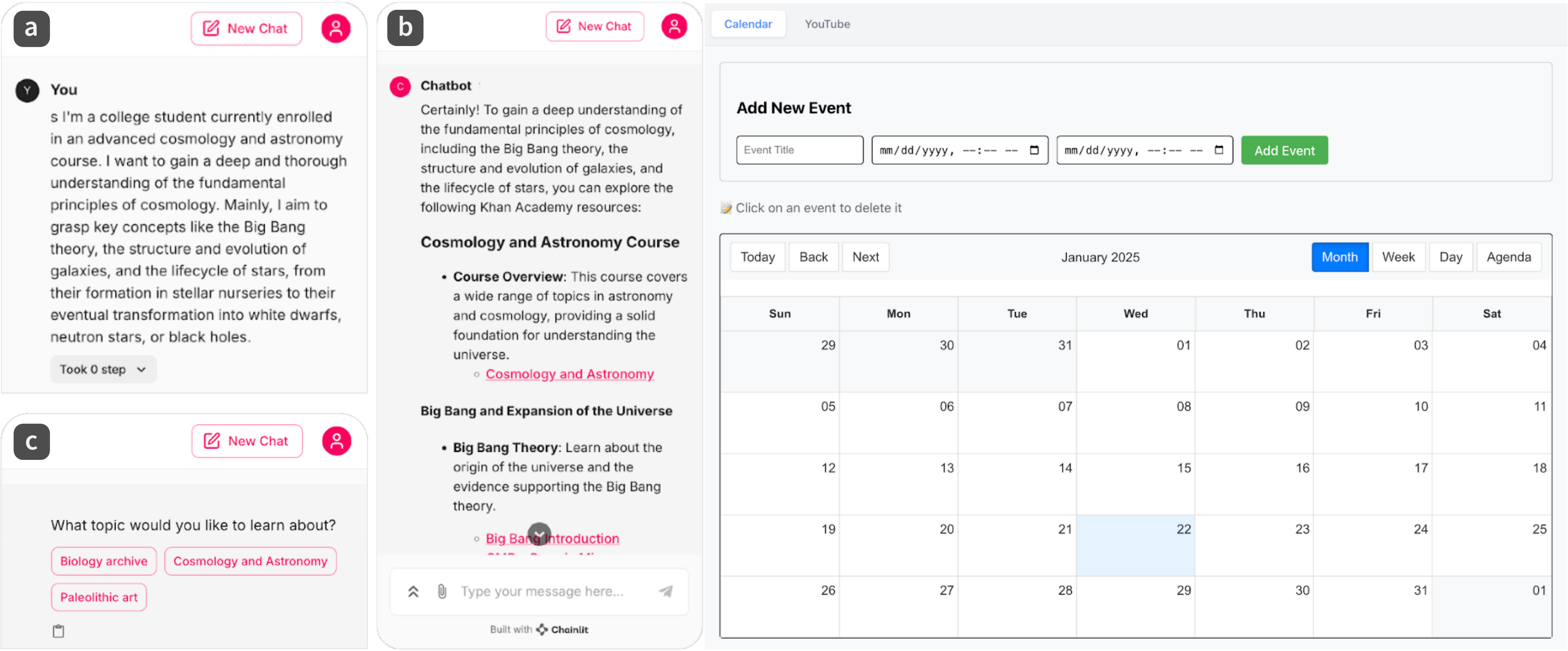}
   \vspace{-12pt}
  \caption{The \ours{} \textit{course selection} interface. Users begin by defining their learning \textit{goals} through either: (1) specifying learning interests (step \protect\darkgreysquare{a}) to generate course recommendations (step \protect\darkgreysquare{b}), or (2) browsing courses by topic (step \protect\darkgreysquare{c}). We outline the user’s interaction with this interface as a guide to explain the pipeline of \ours{} in Section \ref{goalGuidelines}.
}
  \label{fig:courseSelection}
\end{figure*}

  


%% file: figure_personalized_plan.tex
\begin{figure*}[!h]
  \includegraphics[width=\textwidth]{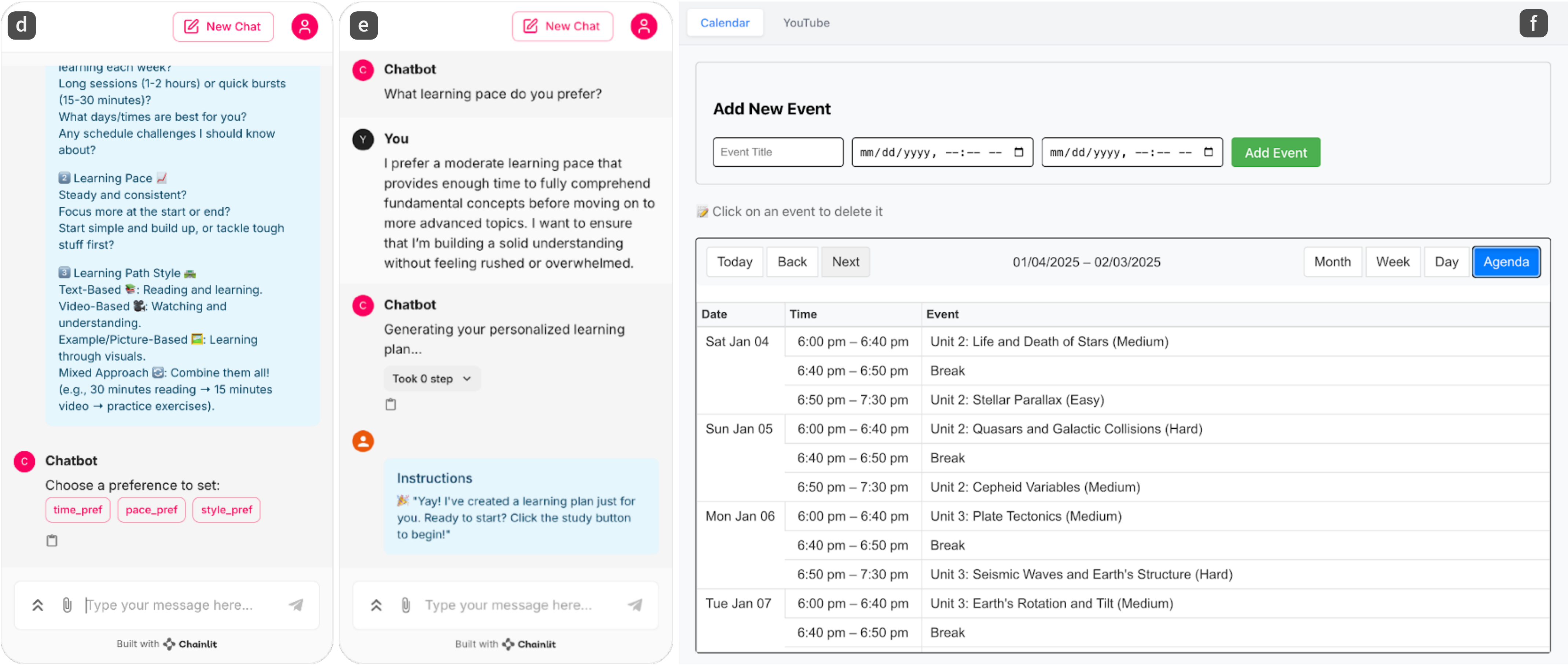}
   \vspace{-12pt}
  \caption{The \ours{} \textit{personalized planning} interface. Users define the remaining dimensions (\textit{time}, \textit{pace}, and \textit{path}) (step \protect\darkgreysquare{d}) in order to generate a learning plan (step \protect\darkgreysquare{e}) and interactive calendar (step \protect\darkgreysquare{f}).
  We outline the user’s interaction with this interface as a guide to explain the pipeline of \ours{} in Section \ref{personalizationGuidelines}.
}
  \label{fig:personalizedPlan}
\end{figure*}

  

%% file: figure_learning_support.tex
\begin{figure*}[!h]
  \includegraphics[width=\textwidth]{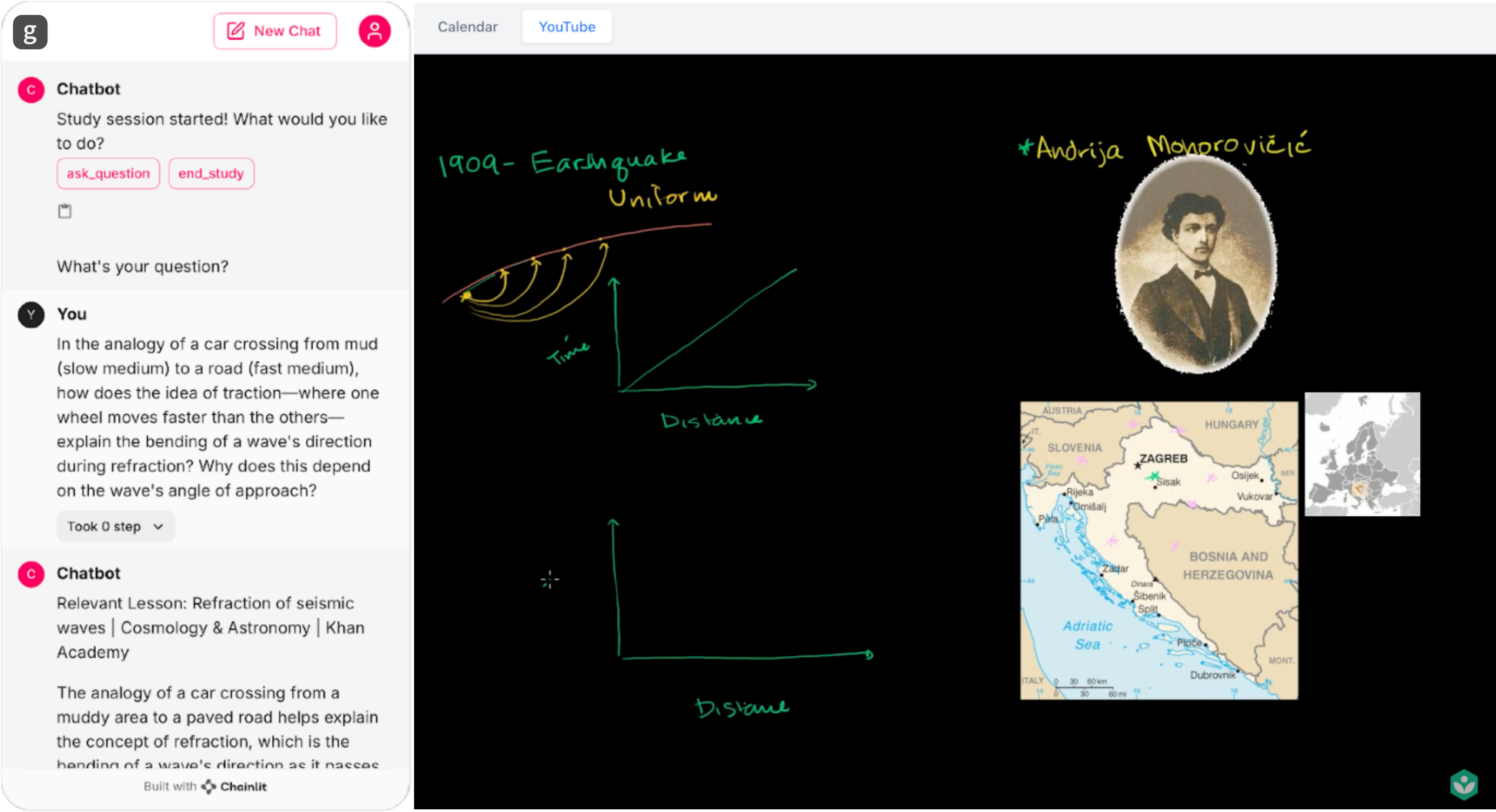}
   \vspace{-12pt}
  \caption{The \ours{} \textit{real-time learning support} interface. Users can ask questions during their learning process and receive immediate, context-specific guidance (step \protect\darkgreysquare{g}).
  We outline the user’s interaction with this interface as a guide to explain the pipeline of \ours{} in Section \ref{RealtimelearningSupport}.
}
  \label{fig:learningSupport}
\end{figure*}